\documentclass{egpubl}
\usepackage{eurova2022}

 \WsPaper           

 \electronicVersion 

\ifpdf \usepackage[pdftex]{graphicx} \pdfcompresslevel=9
\else \usepackage[dvips]{graphicx} \fi

\PrintedOrElectronic

\usepackage{t1enc,dfadobe}

\usepackage{egweblnk}
\usepackage{cite}
\usepackage{subcaption}
\usepackage{color}
\usepackage{cleveref}
\usepackage{ccicons}

\definecolor{c_green}{rgb}{0.01, 0.8, 0.4}

\title[Comparing Nodes of Multivariate Graphs Through Dynamic Layout Adaptations]%
      {Comparing Nodes of Multivariate Graphs\\Through Dynamic Layout Adaptations}
\author[P. Berger, S. Beleites \&  C. Tominski]
{\parbox{\textwidth}{\centering P. Berger~~~S. Beleites~~~C. Tominski\orcid{ 0000-0001-7704-355X}
}
\\
{\parbox{\textwidth}{\centering Institute for Visual \& Analytic Computing, University of Rostock, Germany}
}
}

\begin{document}


\maketitle
\begin{abstract}

Visual comparison is an important task in the analysis of multivariate graphs.
However, comparison of topological features of a graph with respect to its data attributes for different portions of the data remains challenging because there is no single visual representation that would suit the dynamic nature of comparative analyses.
To facilitate the visual comparison in node-link diagrams, we propose the \emph{comparison lens} as a focus+context approach for dynamic layout adaptation.
The core idea is to start with a topology-driven layout and locally inject an attribute-driven layout based on the multivariate similarity of node attributes.
This facilitates comparison tasks on a local level while preserving the user's overall mental map of the graph topology.
Additional visual enhancements, including color-coding, reduction of edge clutter, and radial guides, further support the comparison.
To fit the lens to different comparison situations, it can be configured via user-controllable parameters.
To demonstrate the utility of our approach, we use it for comparison in a real-world dataset of soccer players.


\end{abstract}  
\section{Introduction}

The challenge of analyzing multivariate graphs lies in understanding the graph topology in relation to the graph's multivariate data attributes~\cite{Kerren2014NetworkSurvey}.
In order to find relationships between the graph topology and the underlying attribute values, it is usually necessary to compare different portions of the data with respect to varying analytic goals.
For example, it can be interesting to compare the topology of subgraphs induced by nodes that exhibit some given attribute values.
Or, vice versa, one may want to compare the attribute value distributions of the nodes of certain given graph components.
For a comprehensive understanding of the data, several such visual comparisons need to be conducted.

However, comparison in the context of multivariate graphs is challenging, because it requires visual representations that encode both, the graph topology and the multivariate attributes.
In this work, we concentrate on comparing node attributes when the basic graph visualization is a node-link diagram.
In node-link diagrams, standard topology-driven layout algorithms (e.g., force-directed layout) make the graph topology visible.
One or two selected node attributes can additionally be visualized via on-node encoding~\cite{Nobre2019StateArtVisualizing}.
With this standard approach, comparing nodes regarding their multivariate attributes is difficult.
The reason is that the nodes to be compared could be spatially far apart in the layout, which complicates their comparison, and only very few visualized attributes can be compared directly.

Alternatively, one could use an attribute-driven layout where node coordinates are set based on attribute values. 
This brings similar nodes spatially close to each other, and hence, makes comparisons easier.
However, using an attribute-driven layout usually means that the representation of the graph topology is compromised or even neglected altogether.
Moreover, the graph layout would change globally when different attributes are selected for the layout computation.
This would make it hard or even impossible to develop or maintain a mental map of the overall graph.
Therefore, neither topology-driven layouts with on-node encoding nor attribute-driven layouts alone are ideal for node comparison tasks.

Our goal is to support the flexible comparison of multivariate nodes while keeping the overall graph topology in a node-link diagram intact.
To this end, some form of hybrid layout is required where topological aspects and multivariate attribute values can co-exist, not globally but for a dynamically changing focus of interest.
Inspired by previous work on dynamic layout adaptations, we propose the \emph{comparison lens} as a lightweight interactive tool that can induce transient layout changes into a node-link diagram to make comparison tasks easier.
Within the lens, nodes are laid out according to their multivariate similarity with respect to a selected focus node, which makes their visual comparison straightforward.
Thanks to the focus+context nature of our lens, the overall topology-driven layout is changed only locally so that the cognitive load is reduced and the overall mental map easier to preserve.

After a brief review of related work in \Cref{sec:related}, we describe the functionality of the comparison lens, additional visual enhancements, and user-controllable parameters in detail in \Cref{sec:approach}. We demonstrate the utility of our technique for the case of comparing nodes in a network of soccer players in \Cref{sec:walkthrough}. This paper closes with a brief summary and some ideas for future work in \Cref{sec:concl}.

\section{Related Work}
\label{sec:related}

Our work is related to previous research in multivariate graph visualization, visual comparison, and dynamic adaptation of visualizations.

\paragraph*{Visualizing Multivariate Graphs} The visualization of multivariate graphs is comprehensively discussed in the literature~\cite{Kerren2014NetworkSurvey,Nobre2019StateArtVisualizing}.
Typically, general graph visualization techniques such as node-link diagrams or matrix representations are extended to show the multivariate attributes of the graph, for example via incorporating additional views~\cite{Kerzner2017GraffinityVisualizingConnectivity, Nobre2019JuniperTree+TableApproach}, embedding additional encodings~\cite{Major2019GraphicleExploringUnits, Berger2019VisuallyExploringRelations}, or laying out the graph based on its attributes~\cite{Shneiderman2006SemanticSubstrates, Wu2008VisualizingMultivariateNetworks}.

In this work, we focus on node-link diagrams as a common graph representation.
The literature distinguishes topology-driven and attribute-driven layouts for placing nodes in node-link diagrams.
Topology-driven layout algorithms, such as the force-directed layout~\cite{Fruchterman1991GraphDrawing}, aim to make the graph topology visible.
Attributes can then be represented via additional encodings on the graph nodes~\cite{Dunne2013Motif, Elzen2014MultivariateNetworkExploration, Junker2006OnNodeEncoding}.
On the other hand, attribute-driven algorithms lay out nodes according to their attribute values, for example, as groups in shared regions~\cite{Shneiderman2006SemanticSubstrates, Rodrigues2011GroupBox}, in diagram-like coordinate systems~\cite{Bezerianos2010GraphDice, Eichner2016DirectVisualEditing}, or via multi-dimensional scaling~\cite{Doerk11EdgeMaps}.
As already mentioned, topology-driven and attribute-driven layouts have pros and cons, but comparison tasks remain difficult, in particular when properties of the graph topology need to be interpreted in relation to properties of attributes.

\paragraph*{Visual Comparison}

For visual comparison in general, Gleicher et al.~\cite{Gleicher2011VisualComparisonInformation} propose three fundamental approaches: juxtaposition, superposition, and direct encoding.
Superposition is about showing the elements to be compared on top of each other, juxtaposition creates side-by-side layouts, and direct encoding is based on calculating and visualizing differences directly.
These general approaches are also applied for supporting the visual comparison in graphs, for example, by juxtaposed layouts~\cite{Andrews2009VisualGraphComparison} or superposition of edge information~\cite{Alper2013WeightedGraphComparison}.

Both, juxtaposition and superposition make it easier to carry out comparison tasks.
However, juxtaposed visual representations may conflict with the user's mental map of the graph topology.
For superposition, clutter can be an issue and could lead to misrepresentation of the data.
Direct encoding has the advantage that differences are shown directly.
However, the visualization of differences may be in conflict with the visualization of the graph itself.
Given these different pros and cons, Tominski et al.~\cite{Tominski12VisualComparison} suggest natural interaction techniques allowing users to dynamically create comparison layouts that best suit their task at hand.
Here, we pick up the idea of dynamically adapting a visualization on demand.

\paragraph*{Dynamic Adaptation of Visualizations}

The goal of dynamic visualization adaptations is to provide flexible access to different perspectives on the data while maintaining the general context~\cite{Brueggemann20TheFold, Tominski21FlexibleVA, Horak2021ResponsiveMatrixCells}.
Instead of abrupt switches between views, smooth transitions take users from one perspective to another to keep users in the flow, as proposed by Elmqvist et al.~\cite{Elmqvist2011FluidInteraction}.

Focus+context approaches have long since been used for dynamic adaptation of a region of interest~\cite{Cockburn09FocusContext}.
Interactive lenses are a prominent example of focus+context techniques.
Lens techniques are defined as lightweight tools that induce transient changes into an existing base visualization~\cite{Tominski2017InteractiveLensesVisualization}.

In node-link diagrams, lenses can dynamically adapt the layout to create local neighborhood overviews~\cite{Tominski09CGV} or support navigation via \emph{bring \& go}~\cite{Moscovich2009TopologyAwareNavigation}.
Lenses can also be used to reveal additional details about node attributes~\cite{Jusufi2010NetworkLens}.
Lens-like techniques have also been used to support visual comparison.
For example, the \emph{CompaRing} implements a \emph{bring \& compare} strategy that creates an on-demand juxtaposition of geographic regions to be compared~\cite{Tominski16CompaRing}.

Inspired by these previous works, we propose a lens technique that dynamically adapts the layout of node-link diagrams to support the visual comparison of node attributes in multivariate graphs. To our knowledge, such a dynamic layout adaptation for visual comparison has so far not been described in the visualization literature.

\section{Approach}
\label{sec:approach}

Next, we briefly discuss the requirements of our approach and present the basic idea for a \emph{comparison lens} to facilitate comparison tasks in node-link diagrams.
In addition to the dynamic layout adaptation, we include visual enhancements to further support the comparison and parameters to tune the lens to the situation at hand.

\begin{figure*}
  \includegraphics[width=\linewidth]{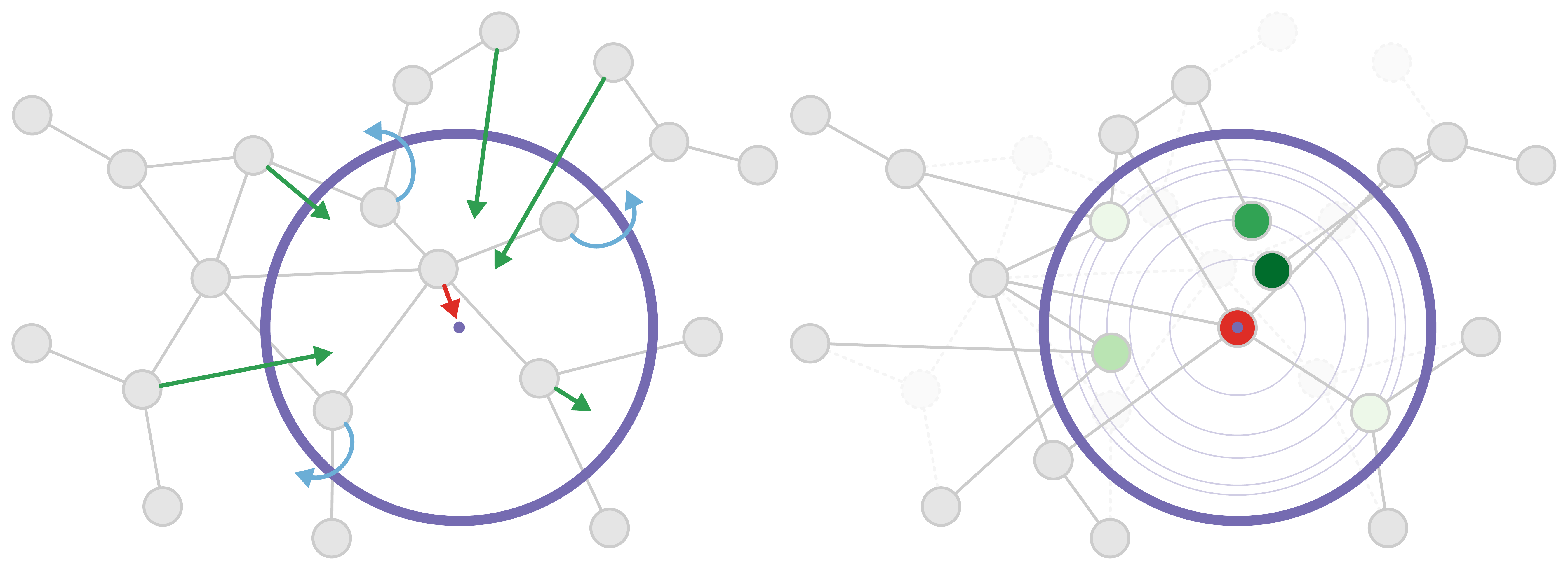}
  \caption{Schematic illustration of the lens-based dynamic layout adaptation (left) and visual enhancements (right) for comparison of nodes. Left: The focus node is centered (red arrow), nodes that meet the similarity threshold are placed according to their similarity with respect to the focus (green arrows), and nodes that do not meet the similarity threshold are moved out of the lens (blue arrows). Right: Different shades of green indicate attribute-based node similarity and radial guides are embedded in the lens to further support the visual comparison.}
  \label{fig:concept}
\end{figure*}

\subsection{Requirements}

Based on the problem description and the review of related work, we identified the following requirements for our approach:

\begin{description}
    \item[R1 Support Visual Comparison:] Given an existing node-link diagram, it should be possible to effortlessly compare selected nodes according to their attribute values.
    \item[R2 Preserve Mental Map:] The user's mental map of the global graph topology should be preserved by reducing global layout changes to a necessary minimum.
    \item[R3 Smooth display transitions:] To keep users in the flow, smooth display transitions should be favored over disruptive view switches.
\end{description}

\subsection{Approach Overview}

The general idea of our approach is to locally induce an attribute-driven layout into an otherwise topology-driven layout of a multivariate graph.
Accordingly, we start with a force-directed node-link diagram that represents the graph topology in a zoomable view. Node size and color can optionally be used to visualize selected node attributes.

To facilitate the comparison of nodes and their attributes in relation to the graph topology as demanded by \textbf{R1}, we propose a lens technique that transforms the graph layout locally based on node attribute values.
Addressing \textbf{R2}, the attribute-driven layout change is restricted locally to the lens interior so that the topology-driven layout is mostly preserved.
To satisfy \textbf{R3}, nodes that are relevant for a comparison task are pulled into the lens and irrelevant nodes are pushed out of the lens via smooth transitions.

Next, we describe how exactly nodes are placed in the lens and how additional visual enhancements and controllable lens parameters further support the comparison.

\subsection{Lens-based Dynamic Layout Adaptation}

The lens is defined by a focal point and a radius around it.
When moving the lens across the node-link diagram, the nearest node is centered within the lens and becomes the focus node, shown in red in Figure~\ref{fig:concept}.
Alternatively, a focus node can be selected using the cursor and the lens is automatically positioned on top of it.

Once a focus node has been specified, the lens interior transforms to an attribute-driven layout that facilitates the comparison of the focus node with other relevant nodes.
Different plausible criteria exist to determine which other nodes should be deemed relevant. One could consider the topological $k$-neighborhood of the focus node or the view neighborhood within a certain window around the lens. As we are primarily interested in node attributes, we use the ``attribute neighborhood'' of the focus node, which boils down to considering the similarity of nodes according to their multivariate attribute values.

To this end, the pairwise attribute-based similarity of the focus node to all other nodes is calculated.
This can be done in different ways, for example, based on average Euclidean distance, Cosine similarity, or Pearson correlation coefficient.
Which nodes are considered relevant is then determined based on a user-adjustable similarity threshold. As indicated by green arrows in Figure~\ref{fig:concept}, nodes that satisfy the threshold, and are thus of interest to the user, are re-positioned inside the lens.
Nodes that are most similar to the focus node are pulled close toward it.
Nodes that just meet the similarity threshold are positioned on the lens boundary.
Moreover, nodes that do not meet the similarity threshold are moved slightly beyond the lens border as shown with blue arrows in Figure~\ref{fig:concept}.
The pulling of relevant nodes toward the lens and the pushing of irrelevant nodes out of the lens can be considered a combination of previously described lens effects from the BringNeighbors lens~\cite{Tominski06Fishey} and the MoleView lens~\cite{Hurter11MoleView}.
The layout adaptation is done in a smooth, force-based animation to make it easy to understand.

In the resulting local attribute-driven layout, the distance of nodes to the lens center represents the similarity of nodes with respect to the focus node.
As all nodes being considered relevant for the comparison are placed inside the lens, they can now be compared based on their distance to the focus node and to each other.
By moving the lens or by selecting another focus node (e.g., from among the re-located nodes), users may continue their comparative exploration of the graph.

While the dynamic layout adaptation already makes comparison tasks easier, it can still require some cognitive effort to compare the distances between nodes.
In the following, we discuss visual enhancements to further support the comparison.

\subsection{Visual Enhancements}

To compare the similarity of two nodes, their corresponding distances to the lens center must be compared.
This can be difficult, if the two nodes to be compared are far apart in the lens.
To address this issue, users can optionally activate additional visual cues to support the comparison.
A first option is to color-code the similarity value on the nodes as show in the right part of Figure~\ref{fig:concept}.

Secondly, concentric circles can be embedded in the lens, acting as radial guides similar to the approach by Moscovich et al.~\cite{Moscovich2009TopologyAwareNavigation}.
Users can select between three different modes: (i) fixed equidistant circles, (ii) fixed circles per node inside the lens, and (iii) a dynamic circle that is linked to the cursor and optionally snaps to nodes.
The radial guides make it much easier to compare nodes with respect to their distance to the center, as can also be seen in Figure~\ref{fig:concept} (right), in particular for the brighter nodes close to the lens border.
The different modes for the radial guides allow users to apply them in a task-specific manner.
Exploratory analysis phases could be supported with a dynamic circle, whereas a few fixed circles can serve confirmatory or presentation purposes.

A final visual enhancement addresses the in-lens clutter of the regular edges of the node-link diagram, which can be severe especially for larger or denser graphs.
To clear potential edge clutter, we combine our lens with an additional lens effect, originally proposed as \emph{LocalEdge} lens by Tominski et al.~\cite{Tominski06Fishey}.
The added effect cuts off edges at the border of the lens if they do not connect to nodes within the lens.
This effectively means that the lens shows only those edges that have an incident node inside the lens. In some situations, it could even be helpful to require both incident nodes of an edge to be within the lens.

The described visual enhancements as well as the dynamic layout adaptation and the lens itself can be fine-tuned via user-controllable parameters, which will be described next.

\subsection{User-controllable Lens Parameters}

Our comparison lens is an interactive tool and as such offers several options for users to control the lens appearance and function.
Most importantly, users can control the focus node, which is closely related to the lens position, the attributes to be used for the similarity calculation, the similarity threshold, the lens size, as well as the mode for the radial guides.

To satisfy changing information needs of users, it is a basic functionality to adjust the lens position via direct-manipulation drag gestures.
As the lens is moved, the node nearest to the lens center is automatically selected as the focus node.
Alternatively, the focus node can be selected via a click, upon which the lens is automatically centered on the new focus node.

To further adapt the lens to different comparison goals, users can interactively specify via a dedicated user interface which node attributes should be taken into account for the similarity calculation.
Users may express their interest in individual attributes and also according to any multivariate combination of attributes.
Changing the attribute selection of the lens results in a re-calculation of all pairwise similarities with respect to the current focus node.
Once the calculation is complete, a smooth display transition updates the node layout inside the lens.

Related to the similarity calculation is the adjustment of the similarity threshold. It determines how many nodes are considered relevant to be affected by the layout adaptation. Currently, the similarity threshold can be set via a slider in the user interface. For a more direct adjustment of this important parameter, it could also make sense to consider on-lens controls as suggested by Kister et al.~\cite{Kister2016MultiLensFluentInteraction}.

Finally, users can adjust the size of the lens through direct manipulation.
This way, it is possible to balance the display space dedicated to the attribute-based visual comparison in the focus inside the lens and the preservation of the graph topology in the context outside the lens.
Moreover, the lens size determines the precision with which the similarity of nodes can be discerned from the attribute-driven layout in the lens interior.
If the comparison task is central to the data analysis a large lens with higher precision can be used.
For brief comparisons, a smaller lens with lower precision may suffice.

In summary, our comparison lens is a novel interactive tool to support on-the-fly visual comparison in multivariate graphs through dynamic layout adaptation and visual enhancements. Several parameters are available to adjust the lens.
Next, we illustrate the utility of our lens for a real-world dataset.

\section{Use Case}
\label{sec:walkthrough}

\begin{figure*}[t]
  \centering
  \begin{subfigure}[b]{0.33\textwidth}
    \includegraphics[width=\textwidth]{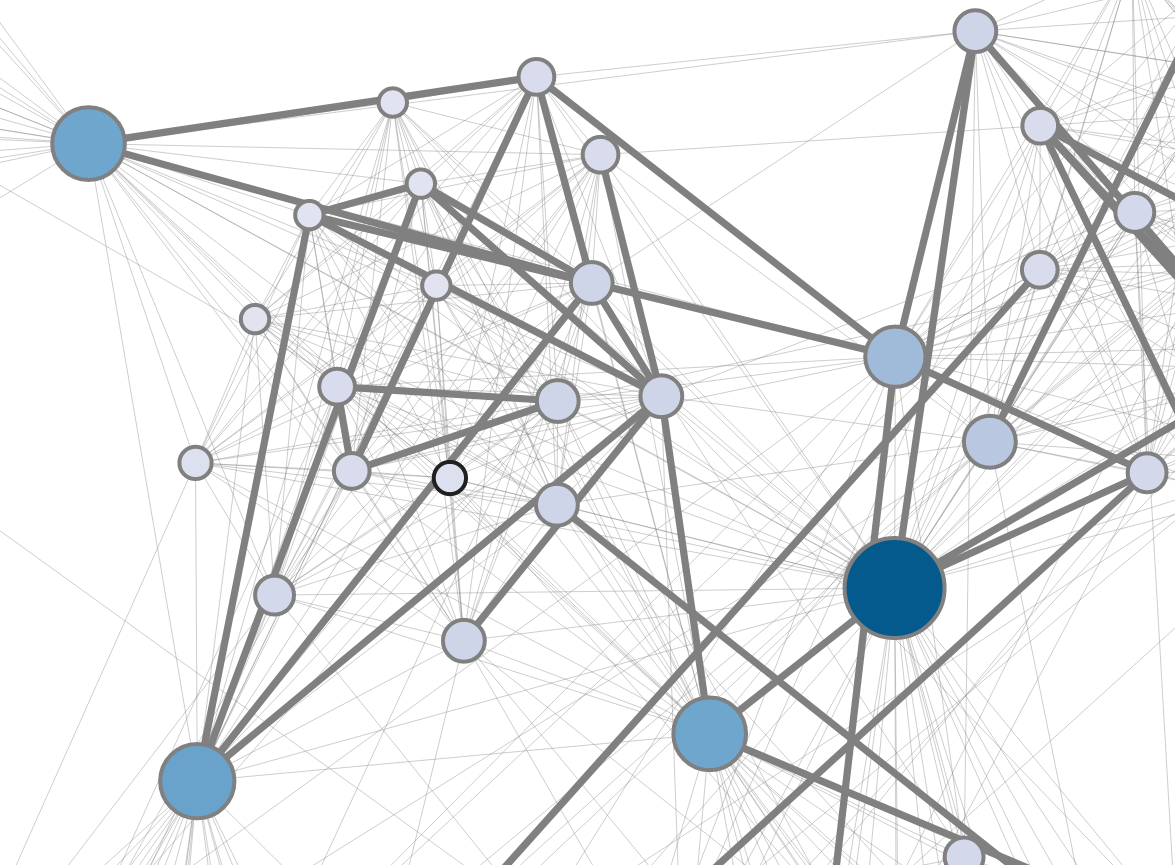}
    \caption{}
    \label{fig:walkthrough1}
  \end{subfigure}
  \hfill
  \begin{subfigure}[b]{0.33\textwidth}
    \includegraphics[width=\textwidth]{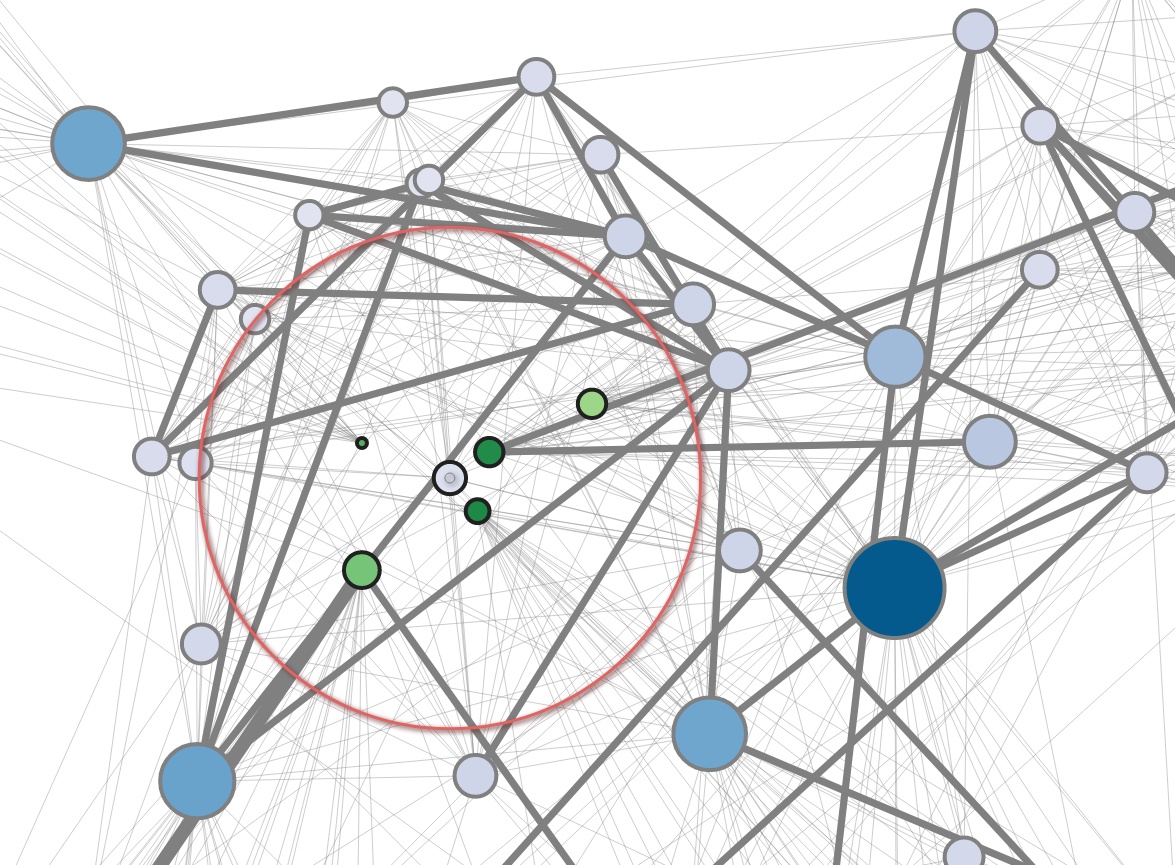}
    \caption{}
    \label{fig:walkthrough2}
  \end{subfigure}
  \hfill
  \begin{subfigure}[b]{0.33\textwidth}
    \includegraphics[width=\textwidth]{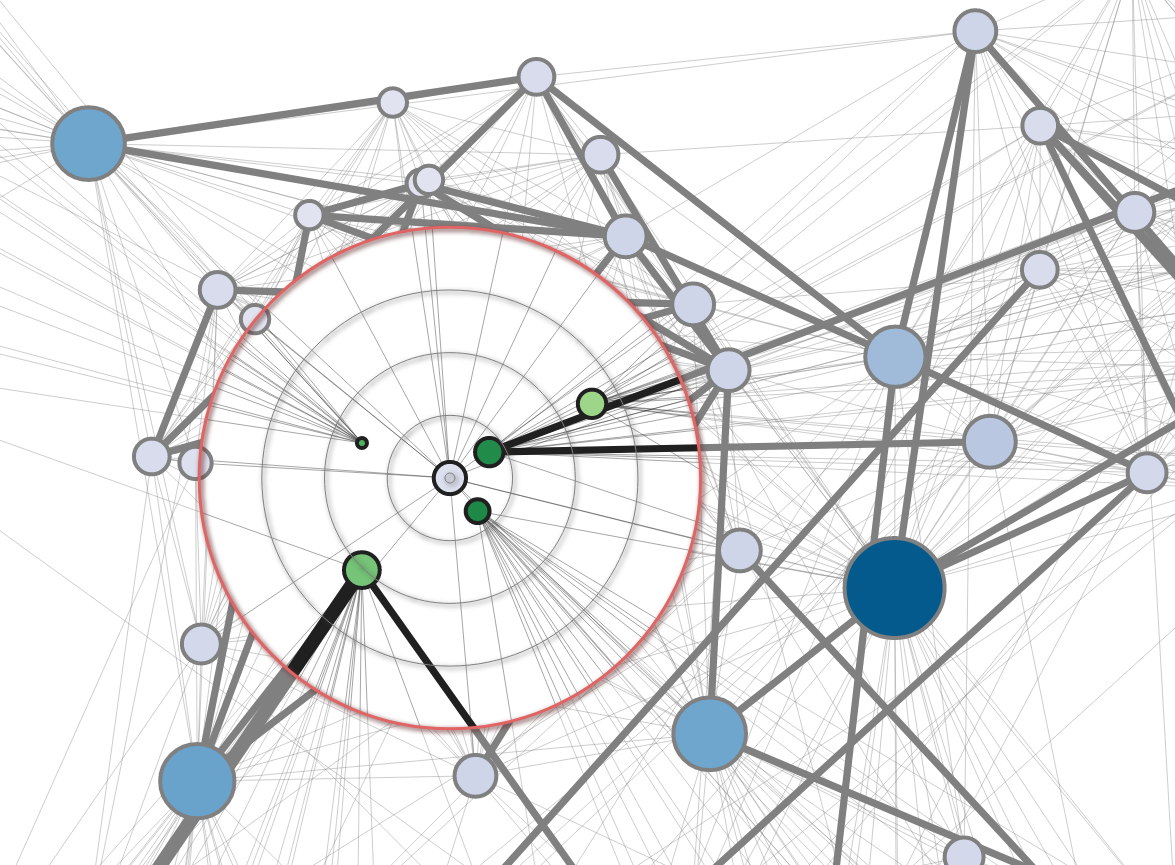}
    \caption{}
    \label{fig:walkthrough3}
  \end{subfigure}
  \caption{Applying the comparison lens to a graph of soccer players. (a) Basic node-link diagram showing the graph topology. (b) Activating the lens pushing irrelevant nodes out of the lens, whereas similar, and hence, relevant nodes are positioned inside the lens based on their similarity. (c) Through edge clutter reduction and radial guides, the comparison inside the lens is enhanced.}
  \label{fig:walkthrough}
\end{figure*}

For a brief demonstration, we use a multivariate graph of soccer players from the Champions League season 2017/18.
Each node of the graph represents a player with multivariate attributes capturing the players' offensive and defensive properties (e.g., minutes played, ball possession, shots on goal, scored goals).
Edges exist between players that have played in the same club.
The number of clubs in which two plays have played is captured as the edge weight.
Overall, the graph has 95 nodes, 1046 edges, and 39 quantitative node attributes.
The basic node-link diagram in Figure~\ref{fig:walkthrough1} shows the graph topology.
Node degree is visualized by node size and shades of blue.
Edge weight is encoded in the link width.

As a team manager, one goal of a comparative data analysis could be to find new players that would fit well in our club.
In our example, we want to find an additional goalkeeper.
For this, we first select our present goalkeeper as the focus node (light blue node with black outline in Figure~\ref{fig:walkthrough1}).
The relevant attributes for goalkeepers to be taken into account for the similarity calculation are set to be \texttt{keeper\_missed} and \texttt{keeper\_save\_total}.

Then we activate the comparison lens in Figure~\ref{fig:walkthrough2}.
Players that do not meet the similarity threshold of 50\% are pushed out of the lens.
Players being sufficiently similar to our selected goalkeeper are moved into the lens and are colored in shades of green to indicate their similarity with respect to the focus node.
As can be seen, five goalkeepers are brought to the lens.
Two of them are positioned particularly close to the lens center, indicating high similarity.
These two players might be interesting as goalkeeper for our team.

For a more detailed look at the situation, we further activate visual enhancements in Figure~\ref{fig:walkthrough3}.
The reduced edge clutter improves the visibility of relevant edges in the lens and makes it easier to investigate the connectivity of similar nodes.
Moreover, the radial guides improve the interpretation of the similarity.
We thus conclude that of the two candidates (dark green nodes), the lower one is slightly more similar and hence should be the player to be approached first for a new contract.

\section{Conclusion}
\label{sec:concl}

Addressing the challenging task of visual comparison in multivariate graphs, we proposed the \emph{comparison lens} as a novel focus+context approach for dynamic layout adaptation to facilitate comparison of nodes in node-link diagrams.
The core idea is to dynamically create a layout where topological features and multivariate attribute characteristics are smoothly intertwined.
Accordingly, nodes are positioned inside the lens with respect to their multivariate similarity, while a topological layout is preserved outside the lens.
This way, cognitive load can be reduced for comparison tasks and the mental map can be preserved easier compared to disruptive view switches.
Additional visual enhancements are incorporated into the lens to further support the comparison.
Users can adapt the lens to their comparison tasks via controllable parameters. 

In the future, we plan to further improve the utility of our approach.
One possible avenue for future work is to enhance the presentation of data attributes within the lens.
Currently, the nodes are color-coded according to their similarity.
However, how individual attributes contribute to the similarity is not visible.
Here it can make sense to dynamically embed additional information.
For example, one could transform the simple dots into multivariate glyphs representing the attributes used for the similarity calculation.

Finally, in order to quantitatively evaluate the utility of our approach, a user study comparing task performance for visual comparison among plain topology-driven layout, an attribute-driven layout, and our dynamic lens-driven hybrid layout would be a worthwhile goal for future work.

\bibliographystyle{eg-alpha-doi}

\bibliography{comparisonlens}

\end{document}